\documentclass[a4paper,11pt]{article}
\pdfoutput=1 

\usepackage{jcappub} 

\usepackage[T1]{fontenc} 

\newcommand{\lya}{Ly$\alpha$}

\newcommand{\mlya}{\mathrm{Ly}\alpha}
\newcommand{\IF}{I-front}
\newcommand{\odrt}{1D-RT}

\newcommand{\mIF}{\mathrm{IF}}
\newcommand{\HI}{H{\sc~i}}
\newcommand{\HII}{H{\sc~ii}}
\newcommand{\HeI}{He{\sc~i}}
\newcommand{\HeII}{He{\sc~ii}}
\newcommand{\HeIII}{He{\sc~iii}}

\newcommand{\appropto}{\mathrel{\vcenter{
  \offinterlineskip\halign{\hfil$##$\cr
    \propto\cr\noalign{\kern2pt}\sim\cr\noalign{\kern-2pt}}}}}

\title{\boldmath Quantifying Lyman-$\alpha$ emissions from reionization fronts}

\author[a]{Bayu Wilson,}
\author[a]{Anson D'Aloisio,}
\author[b]{Christopher Cain,}
\author[c]{Eli Visbal,}
\author[a]{and George D. Becker}

\affiliation[a]{Department of Physics and Astronomy, University of California, Riverside, CA 92521, USA.}
\affiliation[b]{School of Earth and Space exploration, Arizona State University, Tempe, AZ 85281, USA.}
\affiliation[c]{University of Toledo, Department of Physics and Astronomy and Ritter Astrophysical Research Center, Toledo, OH 43606.}

\emailAdd{bwils033@ucr.edu}
\emailAdd{ansond@ucr.edu}
\emailAdd{clcain3@asu.edu}
\emailAdd{elijah.visbal@utoledo.edu}
\emailAdd{georgeb@ucr.edu }

\abstract{
During reionization, intergalactic ionization fronts (\IF s) are sources of \lya\ line radiation  produced by collisional excitation of hydrogen atoms within the fronts. In principle, detecting this emission could provide direct evidence for a reionizing intergalactic medium (IGM). In this paper, we use a suite of high-resolution one-dimensional radiative transfer simulations run on cosmological density fields to quantify the parameter space of I-front Ly$\alpha$ emission. We find that the Ly$\alpha$ production efficiency -- the ratio of emitted Ly$\alpha$ flux to incident ionizing flux driving the front -- depends mainly on the I-front speed and the spectral index of the ionizing radiation. IGM density fluctuations on scales smaller than the typical I-front width produce scatter in the efficiency, but they do not significantly boost its mean value. The Ly$\alpha$ flux emitted by an I-front is largest if 3 conditions are met simultaneously: (1) the incident ionizing
flux is large; (2) the incident spectrum is hard, consisting of more energetic photons; (3) the I-front is traveling through a cosmological over-density, which causes it to propagate more slowly. We present a convenient parameterization of the efficiency in terms of I-front speed and incident spectral index.  We make these results publicly available as an interpolation table and we provide a simple fitting function for a representative ionizing background spectrum.  Our results can be applied as a sub-grid model for I-front Ly$\alpha$ emissions in reionization simulations with spatial and/or temporal resolutions too coarse to resolve I-front structure. In a companion paper, we use our results to explore the possibility of directly imaging Ly$\alpha$ emission around neutral islands during the last phases of reionization. 
}

\begin{document}
\maketitle
\flushbottom

\section{Introduction}
\label{sec:intro}

A detailed understanding of reionization would provide insights into the first ionizing sources in the Universe.  The broad consensus at present is that reionization was likely driven by stellar emissions from high-$z$ galaxies, and that the bulk of reionization was completed over a time spanning $z=5 - 12$. This consensus was built on a combination of observational measurements. Prominent among these are the cosmic microwave background (CMB) constraints on $\tau_{\rm es}$ \cite{2020A&A...641A...6P}, Ly$\alpha$ forest transmission statistics and their spatial fluctuations at $z=5-6$ \cite{2015MNRAS.447.3402B, 2018ApJ...864...53E, 2022MNRAS.514...55B, 2022ApJ...932...76Z}, damping wings in $z > 7$ quasar spectra \cite[e.g.][]{2018ApJ...864..142D, 2020ApJ...896...23W}, and observations of the $z>6$ galaxy population \cite[e.g.][]{Finkelstein_2019, 2021AJ....162...47B, 2023ApJS..265....5H}.  There are extensive ongoing efforts to detect spatial fluctuations in the redshifted 21-cm background, which would constitute the most direct observation of a reionizing IGM.  Such direct evidence of the reionization process remains elusive, however.  

In this paper, and in a companion paper submitted concurrently \cite[][Paper II]{PaperII}, we explore an alternative way to observe the reionization process in action: Ly$\alpha$ emission from cosmological I-fronts.  During reionization, star-forming galaxies carve out ionized patches in the neutral IGM that are bounded by sharp I-fronts which range in width between one to tens of proper kpc's.  The fronts are driven outward at highly supersonic speeds ($10^3-10^4$ km s$^{-1}$) by the ionizing flux from the sources that they enclose \cite{1987ApJ...321L.107S, D_Aloisio_2019, 2019A&A...622A.142D}.   The gas within an I-front is, by definition, a partially ionized mixture of hot electrons and neutral hydrogen atoms. These conditions allow for efficient collisional excitations of the neutrals, which are followed by Lyman series emissions -- the Ly$\alpha$ line being the brightest.  Indeed, such emissions are the chief radiative cooling process limiting the peak temperatures of highly ionized gas left behind by the I-fronts \cite{1994MNRAS.266..343M, D_Aloisio_2019, 2021ApJ...906..124Z}. But they also provide a potential way to detect the I-fronts directly.   A detection of redshifted Ly$\alpha$ structures, ``wispy'' and tens of comoving Mpc across (the sizes of the ionized bubbles or neutral patches bounded by the I-fronts), would be a telltale signature of an ongoing reionization process.  

The possibility of using Ly$\alpha$ emissions from I-fronts to detect reionization has been discussed before. Ref. \cite{Cantalupo_2008} considered emissions from $z\sim 7$ I-fronts driven by bright quasars. Their models predicted \lya\ surface brightnesses of up to $10^{-20}$ erg s$^{-1}$ cm$^{-2}$ arcsec$^{-2}$ at $z=7.1$, which would be detectable with current observatories. Modeling the emissions in more detail, \cite{Davies_2016_IF} later gave a less optimistic forecast. Their calculations showed that quasar I-fronts are likely a factor of $3-15$ dimmer than found by \cite{Cantalupo_2008}.  The difference owes mainly to a more realistic treatment of the effect of cosmological density fluctuations, as well as a relativistic correction to the Ly$\alpha$ intensity -- relevant because quasar I-fronts travel at nearly the speed of light -- first pointed out in \cite{Davies_2016_IF}.   Most recently, \cite{yang2023lymanalpha} used a suite of plane-parallel radiative transfer (RT) simulations to quantify the Ly$\alpha$ intensity and polarization from the more typical galaxy-driven I-fronts during reionization.  In Ref. \cite{koivu2023lymanalpha}, these ``small-scale'' results were applied to a larger-volume semi-numeric simulation of reionization to model the Ly$\alpha$ intensity and polarization power spectra. Their results provide a long-term target for advances in intensity mapping instrumentation.    The work presented here and in Paper II is most similar to that in Refs. \cite{yang2023lymanalpha, koivu2023lymanalpha}, though, as we describe below, our work is motivated by a different observational application, and our results are complementary to theirs in a number of ways. 

Modeling I-front Ly$\alpha$ emissions presents several challenges, including some that go beyond the usual issues facing reionization simulations.   At present, it is prohibitively expensive for state-of-the-art radiative hydrodynamic simulations to sample the large-scale structure of reionization while also resolving the small spatial (sub kpc) and temporal\footnote{To put some numbers to this, a gas parcel will spend less than a Myr in a typical reionization I-front traveling at $v_\mIF=3\times 10^3$ km s$^{-1}$. (Here we have assumed a mean photon energy of 19 eV driving the I-front and that the gas density is the cosmic mean at $z=5.7$.) The collisional excitation cooling timescale for gas at $T=25,000$ K is a few Myr.} scales necessary to model the Ly$\alpha$ emissivity reliably. Moreover, since the Ly$\alpha$ photons scatter many times off of hydrogen atoms as they propagate to the observer, realistic models also need to include the effects of Ly$\alpha$ RT, as in Ref. \cite{yang2023lymanalpha}. One approach to tackling these challenges is to use a suite of idealized high-resolution RT simulations, perhaps in reduced dimensionality, to model the Ly$\alpha$ emissions over the expected parameter space of I-front properties. The results could then be used to map the \lya\ emissivity onto a larger-volume simulation of reionization using the local I-front properties. This was the approach applied in Refs. \cite{yang2023lymanalpha, koivu2023lymanalpha} and we will adopt an approach similar in spirit.

The current paper is complementary to the work of Ref. \cite{yang2023lymanalpha} in a few additional ways: (1) Whereas the simulations of Ref. \cite{yang2023lymanalpha} assume a homogeneous IGM, here we run 1D RT simulations on skewers extracted from cosmological hydrodynamics simulations.  We will compare our main results to a set of uniform-density runs to study the effects of sub-I-front clumpiness on the Ly$\alpha$ emissivity; (2) Ref. \cite{yang2023lymanalpha} adopted a blackbody source spectrum with a range of temperatures.  Here, we adopt a truncated power-law spectrum that is motivated by stellar population synthesis models (see. e.g Ref. \cite{D_Aloisio_2019}).  We believe that our adopted source spectra can be more readily connected to models of the high-$z$ galaxy population in reionization simulations; (3) Lastly, we make our results publicly available in the form of an interpolation table, expressed concisely in a two-dimensional parameter space. We also provide a simple fitting function that will suffice for many applications. To make our results as broadly applicable as possible, we do not include here the effects of Ly$\alpha$ RT, because such effects depend on the setting and application. We will, however, include these effects by way of a Ly$\alpha$ RT code in Paper II, where we model the direct imaging of neutral islands at $z=5.7$.  

The structure of the paper is as follows. In \S \ref{sect:analytic_consideration}, we present simple analytic arguments that provide intuition for our main numerical results. In \S \ref{sect:1DRT} we describe our numerical methodology.  In \S \ref{sect:results} we present our main results and in \S \ref{sect:conclusion} we offer concluding remarks. Throughout this
work, we assume the following cosmological parameters: $\Omega_m$ = 0.3, $\Omega_\Lambda$ = $1-\Omega_m$, $\Omega_b$ = 0.048, and h = 0.68, consistent with the latest constraints \cite{2020A&A...641A...6P}. Distances are given in proper units unless otherwise noted.  For clarity, we will sometimes distinguish between proper and comoving units with a ``p'' and ``c'', respectively, e.g. pkpc or ckpc.

\section{Analytic considerations}
\label{sect:analytic_consideration}
In this section, we present a simple analytic argument to build intuition for the numeric results of the ensuing sections. Here and throughout this paper, what we mean by ``inside of an \IF'' is the region of sharp transition between highly ionized and highly neutral gas bounding a cosmological \HII\ bubble.  There, \lya\ emissions from recombinations are significantly weaker than those of collisional excitations, so to a good approximation we can neglect the former for the discussion in this section (see figure 1 of \cite{Cantalupo_2008}).  The number of Ly$\alpha$ photons from collisional excitations per unit time, per unit volume, emitted at some location within an I-front can be written as,
\begin{align} \label{eq:coll_exc}
    \frac{4 \pi j_{\mlya}}{h_\mathrm{P}\nu_{\mlya}} = n_{\mathrm{e}}n_{\mathrm{HI}} q_{\mlya}^{\mathrm{eff}} (T(r,\alpha)).
\end{align}
Here, $j_{\mlya}$ in [erg s$^{-1}$ cm$^{-3}$ sr$^{-1}$] is the volume \lya\ emission coefficient, $q_{\mlya}^{\mathrm{eff}}$ is the effective collisional excitation coefficient for \lya\ emission [s$^{-1}$ cm$^{-3}$], $n_\mathrm{e}$ and $n_\mathrm{HI}$ are the proper number densities of electrons and neutral hydrogen, respectively, $h_\mathrm{P}$ is Planck's constant, and $\nu_{\mlya}$ is the Ly$\alpha$ frequency. We compute $q_{\mlya}^{\mathrm{eff}}(T)$ using the polynomial fits of Ref. \cite{1987A&AS...70..269G}.

The temperature $T$ depends on $r$, the location within the I-front, as well as the spectral index $\alpha$, which quantifies the shape of the impinging spectrum. Here we assume that the impinging specific intensity of ionizing radiation takes the form $J_\nu \propto \nu^{-\alpha}$. We note that $q_{\mlya}^{\mathrm{eff}}$ is highly sensitive to temperature, with $\Delta T/T \sim 2$ corresponding to $\Delta q/q \sim 10$ at the temperatures of interest ($T_\mathrm{gas}\sim 2-4\times10^4$ K). The \lya\ intensity emitted by an I-front can be found by integrating $j_{\mlya}$ over the width of the front, $R_\mathrm{IF}$. Using the spatial average over the width of the I-front, $\langle j_{\mlya} \rangle = R_\mathrm{IF}^{-1} \int_{R_{\mathrm{IF}}} j_{\mlya}~d r_{\mathrm{IF}}$, the intensity is,
\begin{equation} \label{eq:I_lya_from_jlya}
    I_{\mlya} =   \int_{R_{\mathrm{IF}}} j_{\mlya} dr_{\mathrm{IF}} = 
    \frac{h_\mathrm{P}\nu_{\mlya}}{4\pi} \left< n_{\mathrm{e}}n_{\mathrm{HI}} q_{\mlya}^{\mathrm{eff}} \right> R_{\mathrm{IF}},  
\end{equation}
where $I_{\mlya}$ has units of [erg s$^{-1}$ cm$^{-2}$ sr$^{-1}$]. For the purposes of our simple scaling argument, we will take the thickness of the front to be of order the local mean free path through the neutral gas, $R_{\mathrm{IF}} \sim 1/ \langle n_\mathrm{H} \rangle \bar \sigma_{\nu}$, where $\langle n_\mathrm{H} \rangle$ is the local hydrogen number density averaged over a scale of $\sim 10$ pkpc, characteristic of front widths, and $\bar \sigma_{\nu}$ is the photo-ionization cross-section of \HI\ averaged over the incident spectrum. Ignoring helium for simplicity, we can write $n_{\rm e} n_{\rm HI} = n^2_{\rm H} x_{\rm HI} (1- x_{\rm HI})$, where $x_{\rm HI} \equiv n_{\rm HI}/n_{\rm H}$ is the neutral fraction (which depends on $r$), so that    
\begin{equation}  
    I_{\mlya} \sim   \frac{\langle n_{\rm H}^2 x_{\rm HI} (1-x_{\rm HI})  q_{\mlya}^{\mathrm{eff}} (T) \rangle}{\langle n_{\rm H} \rangle \bar \sigma_{\nu}(\alpha)}. \label{eq:I_lya}
\end{equation} 
The notation $\bar \sigma_\nu(\alpha)$ denotes explicitly the dependence of the average cross section on the shape of the incident ionizing spectrum.  We can write the hydrogen density as $n_{\rm H} = \langle n_{\rm H} \rangle + \langle n_{\rm H} \rangle \delta$, where $\delta \equiv n_{\rm H} / \langle n_{\rm H} \rangle - 1$ is the density contrast with respect to the local mean, $\langle n_{\rm H} \rangle$.  Then eq. (\ref{eq:I_lya}) becomes 
\begin{equation}  
    I_{\mlya} \sim   \frac{ \langle n_{\rm H} \rangle \langle x_{\rm HI} (1-x_{\rm HI})  q_{\mlya}^{\mathrm{eff}} (T) \rangle}{ \bar \sigma_{\nu}(\alpha)} + (\mathrm{terms}~\sim \delta~\mathrm{and}~\delta^2).
    \label{eq:I_lya2}
\end{equation}
Ref. \cite{Davies_2016_IF} pointed out that the boost in $I_{\mlya}$ from IGM clumpiness is subdued because the I-fronts ``resolve'' the density field, i.e. $\delta$ does not deviate far from zero within an I-front.\footnote{Because of the sharpness of I-fronts driven by stellar emissions, this is true even for the cold and clumpy pre-reionization gas.} For the arguments in this section, we will take the first term of eq. (\ref{eq:I_lya2}) to be the dominant one.   

The temperatures inside of I-fronts are determined by the spectrum of the incident ionizing radiation -- here parameterized by $\alpha$ -- and the amount of radiative cooling that occurs inside the front.  The latter is proportional to the amount of the time that the gas spends inside of the I-front, which scales inversely with the velocity of the I-front, $v_\mIF$.  Faster I-fronts will in general have hotter internal temperatures.  For gas overdensities $\lesssim 10$ with respect to the cosmic mean, a reasonable approximation to $T$ may be obtained by considering only $\alpha$ and $v_{\rm IF}$.  However, for larger overdensities $T$ also depends in a non-trivial way on $\delta$ because of the strong dependence of the cooling rate on $T$, so for generality we will write $T \equiv T(\alpha, v_\mIF,\delta)$ (see e.g. the discussion in sections 5.1 and 5.2 of \cite{Davies_2016_IF}).  

The last order of business is to express eq. (\ref{eq:I_lya2}) in terms of $v_{\rm IF}$.  In what follows, we will find it convenient to quantify our results with the ``Ly$\alpha$ production efficiency'', $\zeta$, which we define to be the ratio of the emitted Ly$\alpha$ number flux, $F_{\mlya}^\mathrm{emit}$, to the incident ionizing photon number flux, $F_{\mathrm{LyC}}^{\mathrm{inc}}$. Noting that the I-front speed is\footnote{We assume non-relativistic I-fronts, which is appropriate if stellar emissions dominate the photon budget.} $v_{\mIF} = F_{\mathrm{LyC}}^{\mathrm{inc}}/\langle n_\mathrm{H} \rangle$ (neglecting again the effects of helium), we finally write the Ly$\alpha$ efficiency as,

\begin{equation}\label{eq:lya_efficiency}
    \zeta(\alpha, v_{\mIF},\delta) \equiv \frac{F_{\mlya}^\mathrm{emit}}{F_{\mathrm{LyC}}^{\mathrm{inc}}} \sim \frac{\langle x_{\rm HI} (1-x_{\rm HI}) q_{\mlya}(\alpha, v_{\mIF},\delta)\rangle}{v_{\mIF}\bar{\sigma}_{\nu}(\alpha)}. 
\end{equation}
The goal of this discussion has been to highlight the functional dependence. Note that the explicit dependence on the local density, $\langle n_{\rm H}\rangle$, has dropped out. 
{\it The central takeaway is that we expect the I-front Ly$\alpha$ production efficiency, $\zeta$, to depend on $\alpha$, $v_{\mIF}$, and the clumpiness of the gas on scales within the I-front (encapsulated in $\delta$).} We will see these arguments borne out by our simulation results.  We will find that, while $\delta$ adds scatter to $\zeta$, it does not boost the mean value of $\zeta$ appreciably; on average, $\zeta$ can be parameterized by just two variables: $\alpha$ and $v_{\mIF}$. 

\section{Numeric Methodology}\label{sect:1DRT}
\subsection{One-dimensional Radiative Transfer}

Two of the authors (Cain and Wilson) have developed a one-dimensional radiative transfer (\odrt) code to simulate the propagation of an \IF. We adopt a plane-parallel geometry in which the RT equation takes the simple form,
\begin{equation}\label{eq:td_rt}
dI_\nu = -I_\nu \kappa_\nu dr + j_\nu dr
\end{equation}
where $r$ is the spatial position along the skewer, $I_{\nu}$ is the specific intensity at frequency $\nu$, $\kappa_{\nu}$ is the absorption coefficient of the gas and  $j_{\nu}$ is the emission coefficient.\footnote{It is worth noting that we chose the plane-parallel, rather than spherical, geometry to avoid geometric attenuation of the source ionizing flux over the widths of the I-fronts.  Although such attenuation is small (because the I-fronts are narrow), our preliminary tests showed that it could introduce unwanted effects complicating the interpretation of our main results.}    We discretize the system on a uniform grid with cell size $\Delta r_{\rm cell}$ and place a source of plane-parallel radiation in the first cell that is turned on at time $t_0$.  We adopt a power-law source spectrum, $j_\nu \propto \nu^{-\alpha_{\rm source}}$, between 1 and 4 Ry, and zero at higher energies, which is expected to be a reasonable approximation to the time-averaged ionizing spectrum of metal-poor stellar populations \cite{D_Aloisio_2019}. In general we allow for a time-dependent source emissivity. (We will say more about the usefulness of this in the next section.)  With this boundary condition, the solution to the RT equation is,
\begin{equation}\label{eq:time-dependent_intensity}
    I_\nu(r,t) = I_\nu(0,t)\exp\left[ - {\int_{r_0}^{r} \kappa_\nu(r)dr}\right],
\end{equation}
which we solve numerically using trapezoidal integration.

We track the evolution of \HI, \HII, \HeI, \HeII, \HeIII, and free electrons. The number densities of H and He evolve according to the following equations,
\begin{align}
    \frac{dn_\mathrm{HII}}{dt} &= n_\mathrm{HI} \Gamma_\mathrm{HI} -n_\mathrm{e} n_\mathrm{HII} \alpha_\mathrm{HII}(T) \\
    \begin{split}
        \frac{d n_\mathrm{HeII}}{dt} &= n_\mathrm{HeI} \Gamma_\mathrm{HeI} - n_\mathrm{e} n_\mathrm{HeII} \alpha_\mathrm{HeII}(T) \\ 
        & - n_\mathrm{HeII} \Gamma_\mathrm{HeII} + n_\mathrm{e} n_\mathrm{HeIII} \alpha_\mathrm{HeIII}(T) 
    \end{split} \\
    \frac{d n_\mathrm{HeIII}}{dt} &= n_\mathrm{HeII} \Gamma_\mathrm{HeII} - n_\mathrm{e} n_\mathrm{HeIII} \alpha_\mathrm{HeIII}(T)
\end{align}
where the $\Gamma_{j}$ is the ionization rate (both photoionization and collisional) of species $j$ and $\alpha_{j}$ is the corresponding case A recombination coefficient. 

Whichever, among \{$n_\mathrm{HI}, n_\mathrm{HII}$\} (for H) and \{$n_\mathrm{HeI}, n_\mathrm{HeII}, n_\mathrm{HeIII}$\} (for He), has the smallest abundance is solved for using a first-order implicit backwards-difference scheme. Then the other species are found using the ionizing balance equations,
\begin{align}
    n_\mathrm{H} &= n_\mathrm{HI} + n_\mathrm{HII} \\ 
    n_\mathrm{He} &= n_\mathrm{HeI} + n_\mathrm{HeII} + n_\mathrm{HeIII} \\
    n_\mathrm{e} &= n_\mathrm{HII} + n_\mathrm{HeII} + 2 n_\mathrm{HeIII}
\end{align}
where $n_\mathrm{e}$ is the number density of free electrons. This guarantees that small fractional errors in one quantity do not blow up into huge errors in the other.
The photoionization rate for species $j \in$ \{\HI, \HeI, \HeII\} is given by,
\begin{equation}\label{eq:gamma_species}
    \Gamma^\mathrm{PI}_{j} = \int_{\nu_j}^{\nu_\mathrm{max}}d\nu'\sigma^\mathrm{PI}_{j}(\nu')\frac{4\pi I_{\nu'}}{h_\mathrm{P}\nu'}
\end{equation}
where $h_\mathrm{P}$ is Planck's constant, $h_\mathrm{P}\nu_j$ is the ionization potential of species $j$ (e.g. $h_\mathrm{P}\nu_\mathrm{HI}=1$ Ry), and  $h_\mathrm{P}\nu_\mathrm{max}=4$ Ry.  For the photoionization cross-sections, $\sigma_j^\mathrm{PI}(\nu)$, we use the fitting formulae of \cite{1996ApJ...465..487V}. The collisional ionization rate for each species is given by,
\begin{equation}\label{eq:gammaCI}
    \Gamma^\mathrm{CI}_j = \mathrm{CI}_j (T)n_e
\end{equation}
where CI$_j$ is the collisional ionization coefficient for species $j$, which we obtain from Ref. \cite{Hui_1997}. We do not track secondary ionizations since our source spectra are truncated above $h\nu = 4$ Ry. The total ionization rate is then $\Gamma_j = \Gamma^\mathrm{PI}_{j}+ \Gamma^\mathrm{CI}_j$. 

Temperature evolution is determined by,
\begin{equation}
    \frac{dT}{dt} = \frac{2}{3 k_\mathrm{B} n_\mathrm{tot}}(\mathcal{H}-\Lambda) - 2H(z) - \frac{T}{n_e}\frac{dn_e}{dt}
\end{equation}
where $k_\mathrm{B}$ is the Boltzmann constant, $\mathcal{H}$ and $\Lambda$ are the net heating and cooling rates, respectively, and $H(z)$ is the Hubble parameter. The dominant source of heating is photoionization, 
\begin{equation}
    \mathcal{H}^\mathrm{PI}_j = \int_{\nu_j}^{\nu_\mathrm{max}}d\nu \sigma^\mathrm{PI}_j(h_\mathrm{P}\nu-h_\mathrm{P}\nu_j) \frac{4 \pi I_\nu}{h_\mathrm{P}\nu},
\end{equation}
where $j$ again denotes the species. The net cooling rate consists of collisional excitation, collisional ionization, recombination and inverse Compton cooling:  
\begin{align}
    \Lambda^\mathrm{cooling} &= \Lambda_j^\mathrm{CE}+\Lambda_j^\mathrm{CI}+\Lambda^\mathrm{RC}_j+\Lambda^\mathrm{compton} \\
    \Lambda_j^\mathrm{CI} &= \mathrm{CI}_j(T)n_en_j \\
    \Lambda^\mathrm{RC}_j &=\mathrm{RC}_j(T)n_en_j \\ 
    \Lambda^\mathrm{Compton} &= C(z)n_e(T-T^\mathrm{CMB}).
\end{align}
For the collisional excitation cooling rate, $\Lambda_j^\mathrm{CE}$, we use the fit from Ref. \cite{1992ApJS...78..341C}.  The collisional ionization rate, CI$_j(T)$, case A recombination rate, RC$^\mathrm{A}_j(T)$, and Compton cooling coefficient, $C(z)$, are taken from Ref. \cite{Hui_1997}. Where applicable, the cooling rates adopted here assume gas of primordial composition. Since the redshift remains fixed during our runs, we do not include cooling from cosmic expansion.

The ionization state of the species determines the absorption coefficient in a given cell,
\begin{equation}
    \kappa_\nu =\sum_j n_j \sigma^\mathrm{PI}_j
\end{equation}
where $n_j$ is the number density of species $j$.  The \odrt\ time-step is chosen as the minimum timescale to resolve the cell crossing time, chemical evolution, and thermal evolution. Specifically, the time-step is calculated as the minimum of $\Delta t = 0.1 \frac{x}{dx/dt}$ where $x \in$ \{$n_\mathrm{HI}, n_\mathrm{HII}, n_\mathrm{HeI}, n_\mathrm{HeII}, n_\mathrm{HeIII},T$\} as well as the cell crossing time, $\Delta t = \Delta r_\mathrm{cell} / c$, where $c$ is the speed of light. We have confirmed that the physical quantities of interest are converged with this method (see Appendix \ref{App:convergence}).

\subsection{Simulating I-front Ly$\alpha$ emissions}
\label{sect:sim_lya_emission}

We ran a suite of RT simulations to quantify $\zeta$ over a wide range of $\alpha$ and $v_{\rm IF}$. To understand the effect of density fluctuations on the Ly$\alpha$ emissions, we have performed two sets of runs. One set adopts a uniform density field and the other uses skewers traced through a cosmological hydrodynamics (hydro) simulation, hereafter our ``fluctuating-density'' runs.  For the latter set, we employ the hydro simulation used in Ref. \cite{Davies_2016_IF}, which was run with the GADGET-3 code \cite{2005MNRAS.364.1105S}. The simulation box has a side $L= 3~h^{-1}$Mpc with $2\times 512^3$ dark matter and gas particles. The gas was not photoheated by an ionizing background\footnote{However, a temperature floor of 50 K was imposed.}, making the simulation useful for illustrating the upper range of effects that density fluctuations can have on I-front Ly$\alpha$ emissions. 

We have been given access to only the $z = 7.1$ snapshot of the hydro simulation. From this box, we extract thirty 1.5 pMpc-long skewers.  We select only skewers with $n_\mathrm{H}<10^{-2} $cm$^{-3}$ everywhere in order to exclude self-shielding density peaks, which would otherwise trap the I-front that we send along the skewer because we do not include the hydrodynamic reaction from photoheating in our runs.  This threshold corresponds roughly to a gas overdensity of $\sim100$ times the cosmic mean density at redshift $z\sim 5.7$, i.e. near the end of reionization. The densities along the skewers were interpolated onto a uniform grid with cell size $\Delta r_{\rm cell} = 0.25$ pkpc, while for the uniform density runs we use $\Delta r_{\rm cell} = 0.50$ pkpc.  In Appendix \ref{App:convergence}, we show that in both sets of runs our results are converged with respect to the cell size. 

Since our focus in Paper II is modeling Ly$\alpha$ emissions from neutral islands, we rescale the hydro skewer densities to a redshift of $z=5.7$.\footnote{For reference, the mean gas density of the re-scaled skewers is $\bar \rho_\mathrm{gas}=1.3\times10^{-28}$ g cm$^{-3}$ which corresponds to a hydrogen number density of $n_\mathrm{H}=5.7\times10^{-5}$  cm$^{-3}$.} This is our fiducial redshift because it is approximately the central redshift of the 110 $h^{-1}$cMpc-long Ly$\alpha$ forest trough found in the spectrum of QSO J0148 \citep{2015MNRAS.447.3402B}. 
 It is also the central redshift of narrow band filters commonly used to search for Ly$\alpha$ at $z=5.7$.  The NB816 filter on the Subaru Telescope Hyper Suprime-Cam, for example, has been used to map Ly$\alpha$-emitting galaxies around the J0148 trough \citep{2018ApJ...863...92B, 2021ApJ...923...87C, 2023ApJ...955..138C}. Searching near the J0148 trough at $z=5.7$ would be a natural place to start looking for the signal explored here. We emphasize, however, that our results are relatively insensitive to redshift (and therefore cosmic mean density). For example, we ran two test cases on uniform-density skewers at $z=5$ and $z=6.6$, for which the densities differ by factor of 2, and found that the \IF-averaged temperature and \lya\ photon production efficiency only differed by $<3\%$. 
 
The RT source is turned on in a fully neutral IGM. The gas does not evolve dynamically and cosmological expansion is turned off during the runs. For both sets of runs, we use 7 values of $\alpha$ equally spaced between $0 \leq \alpha\leq 3$, and our runs are designed to span $ 300 \lesssim v_{\rm IF} \lesssim 5 \times 10^4$ km/s, representative of I-front speeds driven by stellar emissions during reionization (see e.g. \cite{D_Aloisio_2019}).  We implement two features in our setup that were designed to give us full control over $\alpha$ and $v_{\rm IF}$ in our \odrt\ runs: 
 
 \begin{enumerate}
 \item 
 We have argued that the Ly$\alpha$ production efficiency, $\zeta$, should depend on the spectral index, $\alpha$, of the ionizing radiation that is incident on the I-front.  Note, however, that $\alpha$ is generally not the same as the spectral index of the source, $\alpha_{\rm source}$, owing to filtering by H ionizations along the line of sight.  The longer the path length from source to I-front, the more spectral hardening occurs, especially if the radiation encounters strong over-densities. When quoting our results, we would like to remove this dependence on the path-length to the source in order to make them more general and easier to apply to any I-front, e.g. in reionization simulations outside of this work.  We take the approach of Ref. \cite{D_Aloisio_2019} to effectively remove spectral hardening by sharply increasing the photoionization rate (artifically) as soon as the gas reaches ionization equilibrium. In practice, we follow the condition used in Ref. \cite{Davies_2016_IF} to find the equilibrium cells: 
\begin{equation}
     \frac{\dot{n}_\mathrm{e}}{n_\mathrm{tot}}\frac{\Delta r_\mathrm{cell}}{c} <10^{-8}
\end{equation}
where $\dot{n}_\mathrm{e}$ is the rate at which the electron number density changes with time, $n_\mathrm{tot}$ it the total number density of hydrogen, helium, and electrons, and $c$ is the speed of light. We enforce that cells where this condition holds contribute essentially no opacity along the sight line.  This ensures that $\alpha = \alpha_{\rm source}$ to a good approximation, which allows us to control $\alpha$ in our ``numerical experiments''. The lack of attenuation behind the I-front also allows us to read off the incident ionizing flux, $F^{\rm inc}_{\rm LyC}$, trivially from the source flux. (Recall that $F^{\rm inc}_{\rm LyC}$ is needed to obtain the efficiency $\zeta$, eq. \ref{eq:lya_efficiency}.)      We have checked that this procedure does not spoil the I-front structure by running a series of tests comparing \IF\ position and speed from \odrt\ with analytic solutions with and without temperature evolution. We found that, in any case, there is never more than a 5\% difference between the analytic and measured quantities. 

\item I-front speeds depends on both the flux of ionizing radiation and local density.  For a fixed source luminosity, our set of uniform-density runs would not sample the desired range of I-front speeds because of the lack of geometric attenuation in the plane-parallel geometry.   To efficiently sweep a wide range of $v_{\rm IF}$ in our uniform-density runs, we implement a time-dependent emission coefficient at the source cell that decays as $t^{-2}$. (Our results are independent of the exact form of the time-dependence; its sole purpose is to explore a range of $v_{\rm IF}$.)  For our fluctuating-density runs, i.e. on the hydro simulation skewers, density fluctuations cause \IF\ speed variation. In these runs, we also assign each skewer a constant source luminosity drawn from a distribution that samples the desired range of \IF\ speeds.

\end{enumerate}
\begin{figure}
    \centering
    \includegraphics[width=0.6\textwidth]{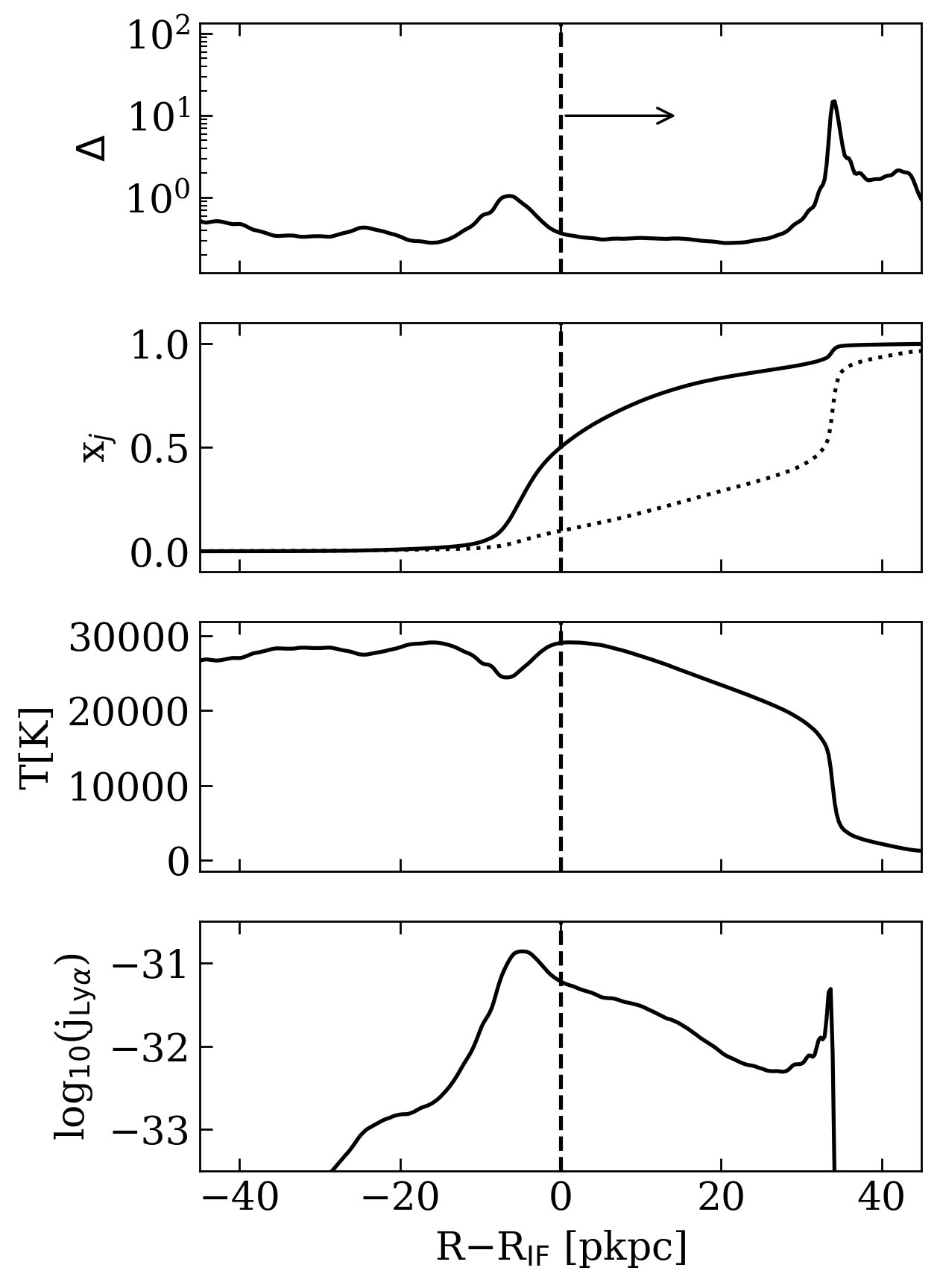}
    \caption{A zoom-in on a typical \IF\ region traveling through a fluctuating-density skewer with incident spectral index $\alpha = 1.5$. The dashed line denotes the \IF\ location, marked where $x_\mathrm{HII}=0.5$, and the arrow indicates the direction that the \IF\ propagates. Going from top to bottom, we display the mass density field, in units of the cosmic mean, the neutral fraction for \HI\ (solid) and \HeI\ (dotted), the gas temperature, and the \lya\ emission coefficient, where $j_{\mlya}$ has units of [erg s$^{-1}$ cm$^{-3}$ sr$^{-1}$].}
    \label{fig:profileIF}
\end{figure}

In Figure \ref{fig:profileIF}, we zoom in on a snapshot of a typical \IF\ region from our fluctuating-density runs. The Ly$\alpha$ emission exhibits two peaks as a result of the density structures within the I-front. The center of the \IF\ is slightly in front (rightward) of a slight overdensity, which shortens the local mean-free-path of ionizing photons and slows down the rear of the I-front. The slow-down allows more time for collisional excitation cooling.  Note also that the cooling rate is enhanced at higher density.  These effects result in the broad peak in $j_{\mlya}$ to the left of the vertical dashed line. The front(right)-half of the \IF\ contains a larger overdensity with $\Delta \approx 10$, which similarly results in the sharp peak in $j_{\mlya}$ on the right, as well as the kink in the neutral fractions.  This example highlights that sub-\IF\ density structures play a non-trivial role in setting the \lya\ emissivity. 

\begin{figure*}
    \centering
    \includegraphics[width=1.0\textwidth]{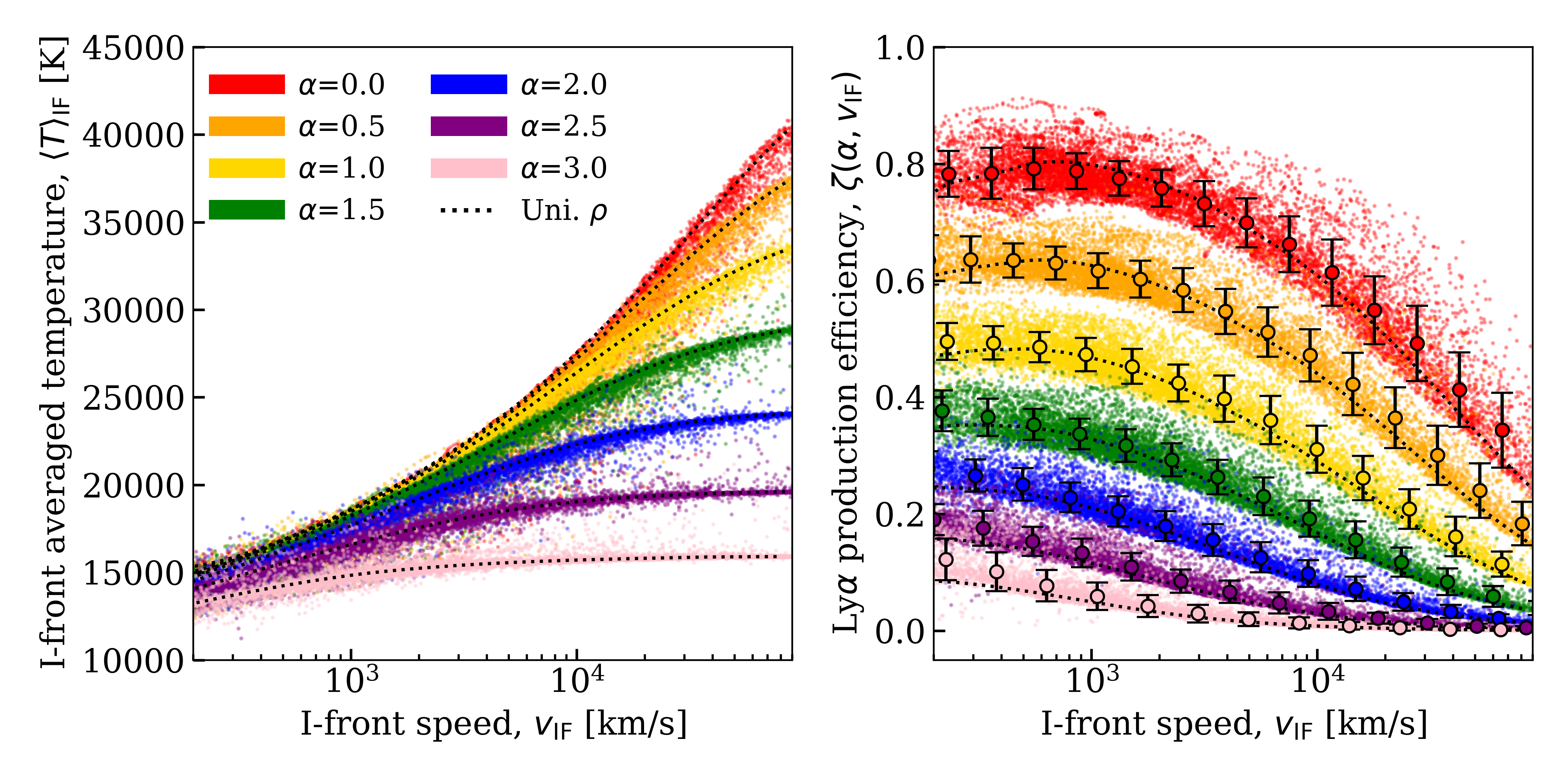}
    \caption{Examples from our \odrt\ runs illustrating the dependence of I-front internal temperature (left) and Ly$\alpha$ production efficiency (right) on I-front speed, $v_{\rm IF}$, for fixed spectral index of incident ionizing radiation, $\alpha$.   The dotted/black curves are from uniform-density runs, while the colored data points are from our fluctuating-density runs. In the right panel, the open circles with error bars give the mean values and standard deviations from the fluctuating-density results, in logarithmically spaced $v_{\rm IF}$ bins. These track reasonably closely the corresponding uniform-density results.  The main effect of small-scale density fluctuations within the I-fronts is to produce scatter. They do not boost the mean Ly$\alpha$ production efficiency appreciably. }
    \label{fig:temperature_lyaEff_plot}
\end{figure*}

Lastly, we address the important question of whether the spectrum of incident ionizing radiation driving an intergalactic I-front can be reasonably approximated by a power law, as we have assumed here. Ref. \cite{D_Aloisio_2019} explored the time-integrated spectra of metal poor stellar populations using stellar population synthesis modeling. They showed that a power-law $I_{\rm \nu} \propto \nu^{-\alpha}$ between 1 and 4 Ry is a reasonable approximation to these spectra. However, even if the emitted source spectrum were a perfect power law, H absorptions between the source and I-front could change the shape of the spectrum considerably. In Appendix \ref{app:spectra}, we explore this effect with 1D-RT simulations. Assuming a perfect power-law for the source spectrum, we find significant deviations in the spectrum shape after traveling a distance $\gtrsim 5~h^{-1}$cMpc from the source through the clumpy ionized IGM.  However, the ionizing background incident on any I-front during reionization will be an integrated contribution from many sources at a distribution of distances. We model this integration effect using sources from a cosmological simulation and find more modest shape differences in the source-population averaged spectrum (see right panel of Figure \ref{fig:power_law_tests}). From these tests we conclude that, if a power-law is a reasonable approximation to the intrinsic spectrum of the sources, then a power-law (with hardened spectral index) likely remains a reasonable approximation to the spectral shape of the radiation driving intergalactic I-fronts, in spite of filtering by the intervening IGM.

\section{Results}
\label{sect:results}
In section \ref{sect:analytic_consideration}, we argued that gas temperatures within I-fronts depend on $\alpha$ and $v_{\rm IF}$, which led us to the assertion that the Ly$\alpha$ efficiency, $\zeta$, should scale with these quantities as well.  Figure \ref{fig:temperature_lyaEff_plot} shows this borne out in our simulations.  The left panel illustrates the dependence of I-front-averaged gas temperature on $v_{\rm IF}$ for a number of $\alpha$ values.  The right panel shows the same for $\zeta$. To reiterate, $\zeta$ is a conversion factor between the incident ionizing photon number flux and the emitted Ly$\alpha$ number flux (see eq. \ref{eq:lya_efficiency}). In both panels the colored data points are the results from our fluctuating-density runs, while the dotted-black curves are the results from our uniform-density runs.  In the right panel, the open black circles and error bars correspond to mean values and standard deviations in logarithmically spaced bins of $v_{\rm IF}$ for our fluctuating-density runs.  Here and throughout the rest of this paper, we calculate $v_{\rm IF}$ using,
 \begin{equation}
    \left<v_\mathrm{IF}\right>_\mathrm{ion} = \int_0^1 v_\mathrm{IF} dx_\mathrm{HII} \\
    = \int_\mathrm{IF} \frac{dx_\mathrm{HII}}{dt} dr \rightarrow \sum_{i \in {\rm cells}} \Delta r_{\rm cell}^{i} \frac{\Delta x_{\rm HII}^{i}}{\Delta t}
\end{equation} 
where $x_{\rm HII}$ is the local ionized hydrogen fraction, and the right arrow denotes discretization on the RT grid, with $\Delta r_{\rm cell}^{i}$  and $\Delta t$ being the width of cell $i$ and time step, respectively. Note that the $v_{\rm IF}$ under the first integral is the velocity of a point within the I-front at a fixed $x_{\rm HII}$, while $dx_{\rm HII}/dt$ is evaluated at fixed position.  As such, the second equality involves a transformation from the Lagrangian to Eulerian frame.  For test cases in which the IF is sufficiently well-resolved to evaluate the first expression accurately, we find the two definitions to be equivalent.  The main motivation for using this method of calculating $v_{\rm IF}$ is that it is substantially less noisy and more representative of the whole \IF\ compared to the simpler method of tracking a location of fixed $x_{\rm HI}$ (e.g. $x_{\rm HI}=0.5$), which is sensitive to sub-\IF\ fluctuations. We calculate the \IF\ averaged temperature in a similar way, $\left<T_\mathrm{IF}\right> = \int_0^1 T_\mathrm{IF} dx_\mathrm{HII}$, and for $\zeta$ we integrate over the I-front\footnote{In practice, because the emission is strongly peaked within the front, this integral is not sensitive to the choice of I-front boundaries (for reasonable choices), so we extend the limits of integration to include the whole skewer.} to obtain a Ly$\alpha$ number flux and then divide by the incident ionizing flux.  

Figure \ref{fig:temperature_lyaEff_plot} shows that there is a one-to-one relation between $\zeta$ and $v_{\rm IF}$ for fixed $\alpha$ in our unifom-density runs, and that the fluctuating-density runs follow close to the same curves {\it on average}. This broadly supports the arguments of \S \ref{sect:analytic_consideration} which led us to express $\zeta$ as a function of $\alpha$ and $v_{\rm IF}$. Importantly, we find that the sub-I-front density fluctuations do not substantially boost the Ly$\alpha$ emissions on average, consistent with the results of Ref. \cite{Davies_2016_IF}. They attributed this effect to the fact that the I-fronts are sharp enough to partially ``resolve'' the density field, i.e. variations in density within the small-scales spanned by an I-front are typically mild. We note that the inverse relation between density and local mean free path provides a kind of adaptive ``resolution'' as well.  I-fronts get narrower as they encounter a density peak.   

Figure \ref{fig:temperature_lyaEff_plot} also makes clear that the main effect of sub-I-front density fluctuations is to induce a scatter in $\zeta$. The scatter increases for harder incident ionizing spectra (lower $\alpha$) because the I-fronts are wider for harder spectra, owing to the energy dependence of the photoionization cross section. Overall, the results in Figure \ref{fig:temperature_lyaEff_plot} suggest that the parameter space of the mean Ly$\alpha$ efficiency can, to a good approximation, be covered with just two variables: $\alpha$ and $v_{\rm IF}$.  In other words, modeling the intrinsic Ly$\alpha$ emissions from I-fronts in reionization simulations can be boiled down to the spectrum of incident ionizing radiation and the I-front speed. 

\begin{figure*}
    \centering
    \includegraphics[width=0.48\textwidth]{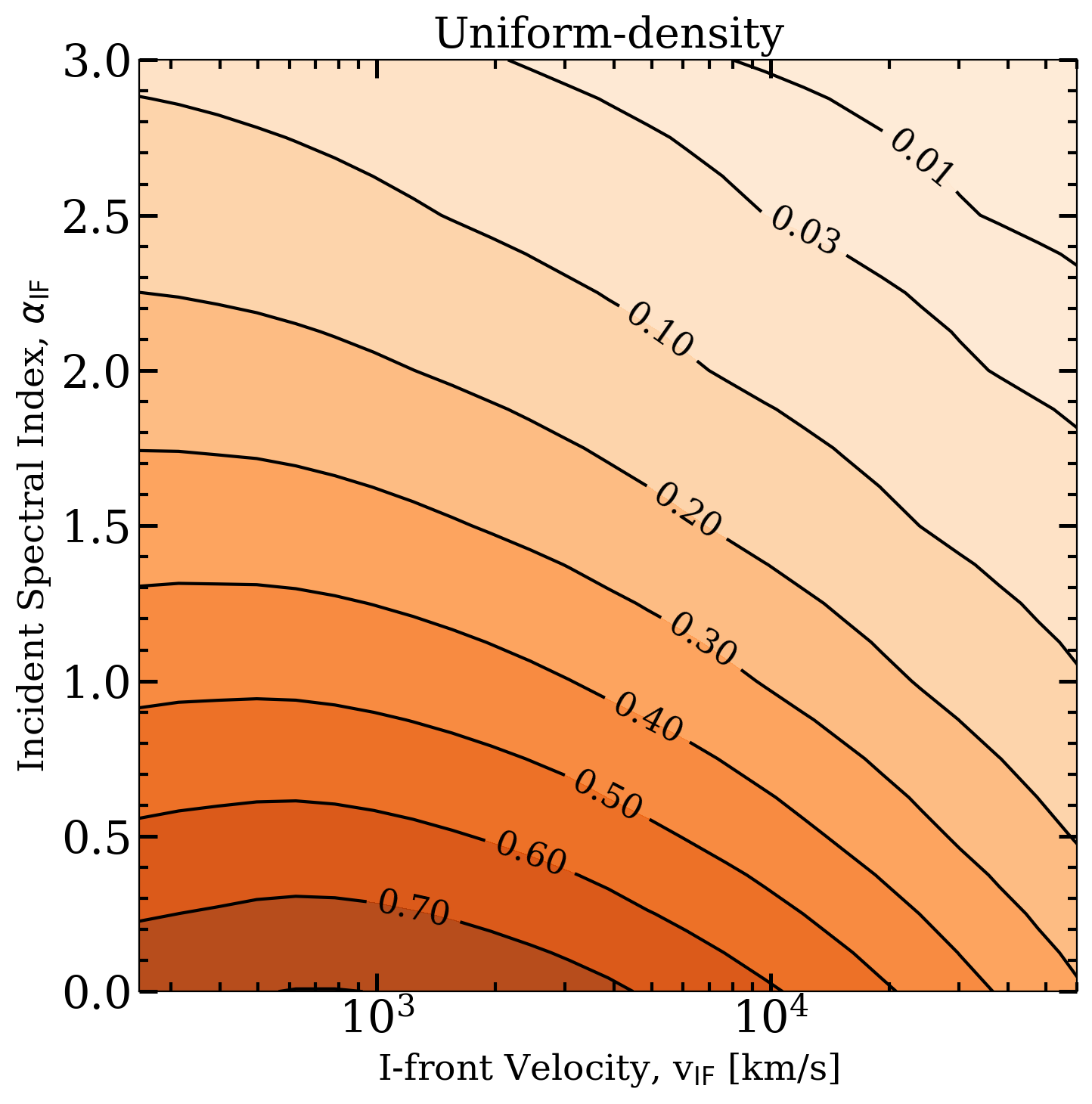}
    \includegraphics[width=0.48\textwidth]{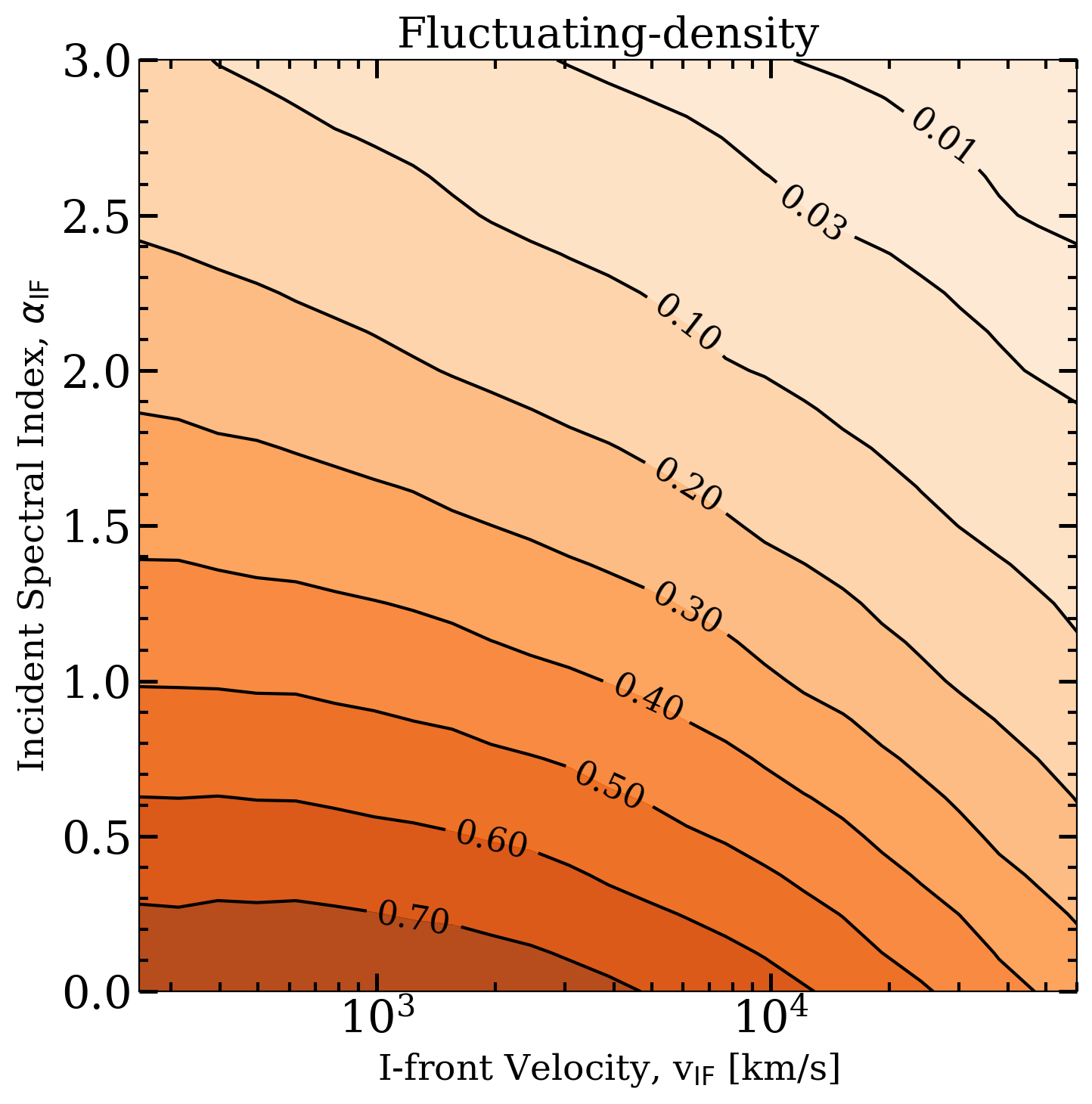}
    \caption{Parameter space of the I-front Ly$\alpha$ production efficiency, $\zeta$, i.e. the ratio of emitted Ly$\alpha$ number flux to the flux of ionizing photons being absorbed by the I-front. The x-axis refers to the I-front speed, $v_{\rm IF}$, while the y-axis refers to the spectral index of ionizing radiation being absorbed by the I-front, $\alpha$. The curves are contours of constant efficiency. The left and right panels correspond to our uniform- and fluctuating-density runs, respectively.   (In the fluctuating-density case, the contours represent constant mean efficiency in logarithmically spaced bins of $v_{\rm IF}$.) The similarity of the left and right panels shows that IGM density fluctuations within I-fronts do not substantially boost the Ly$\alpha$ production efficency on average. These results may be applied as a sub-grid model for I-front Ly$\alpha$ emissions in large-volume reionization simulations that do not resolve the internal structure of I-fronts. An interpolation table for these results may be found here: \href{https://github.com/bayu-wilson/1DRT_quantify_reion_fronts/tree/main/results/interp_tables/fd_parameter_space_interp_table.txt}{https://github.com/bayu-wilson/1DRT\_quantify\_reion\_fronts/tree/main/results/interp\_tables}.     }
    \label{fig:pspace_lya_eff_fd}
\end{figure*} 

The main results of this paper are shown in Figure \ref{fig:pspace_lya_eff_fd}: the parameter space of \lya\ production efficiency for a range of likely $\alpha$ and v$_{\mathrm{IF}}$ values during reionization. The left and right panels correspond to our uniform- and fluctuating-density runs, respectively. The curves are contours of constant efficiency. (For the fluctuating-density case, we show contours of mean efficiency in logarithmically spaced bins of $v_{\rm IF}$. For brevity, we will also refer to the mean value as the efficiency.) First, note the similarity of the uniform- and fluctuating-density results, consistent with our discussion above. Even when cosmological fluctuations are included in our calculations, the mean efficiency tracks closely the results from our uniform-density runs.

In both plots, the efficiency is highest when \IF\ speeds are slower and the incident spectrum is harder (smaller $\alpha$), i.e. toward the lower left corner of the parameter space. Under these conditions, there are more Ly$\alpha$ photons produced per incident ionizing photon. Consider, for example, an I-front traveling at $10^4$ km s$^{-1}$, driven by an incident ionizing radiation background with $\alpha = 1.5$.  Reading values from Figure \ref{fig:pspace_lya_eff_fd} at $(\alpha,v_\mathrm{IF}) = (1.5,10^4$ km s$^{-1})$, we see that the \lya\ efficiency will be $\zeta \sim 0.19$, or approximately 2 Ly$\alpha$ photons will be emitted for every 10 ionizing photons that are absorbed at the back of the I-front.   If this \IF\ were to encounter an over-density that decreased its speed such that $(\alpha,v_\mathrm{IF}) = (1.5,10^3$ km s$^{-1})$, the efficiency would increase by a factor of 1.8 to $\zeta \sim 0.33$. If, elsewhere, an I-front is being driven by a harder incident radiation field such that $(\alpha,v_\mathrm{IF}) = (0.5,10^4$ km s$^{-1})$, the \lya\ efficiency would be $\zeta \sim 0.46$, a factor of $2.5$ larger than the first case.  Lastly, if $(\alpha,v_\mathrm{IF}) = (0.5,10^3$ km s$^{-1})$, the  efficiency is $\zeta \sim 0.62$, a factor of 3.3 larger than the first case.

When interpreting Figure \ref{fig:pspace_lya_eff_fd} in terms of actual Ly$\alpha$ emission, it is important to keep in mind that the quantity we show is not the same as the flux of \lya\ photons emitted by an I-front. Rather, the flux can be obtained by multiplying the efficiency and incident ionizing flux, $F_{\mlya}^\mathrm{emit} = F_{\mathrm{LyC}}^{\mathrm{inc}}~\zeta(\alpha, v_{\mIF})$. Recall also that $F_{\mathrm{LyC}}^{\mathrm{inc}}$ is related to $v_{\rm IF}$ by $F_{\mathrm{LyC}}^{\mathrm{inc}} = v_{\rm IF} \langle n_{\rm H}\rangle$.  As an illustrative example, consider two scenarios with fixed incident spectral index, $\alpha$, and fixed density, $\langle n_{\rm H} \rangle$. One I-front is moving at $v_{\rm IF} = 10^3$ km s$^{-1}$ and the other at $v_{\rm IF} = 10^4$ km s$^{-1}$, which implies that the latter is being driven by ten times the $F_{\mathrm{LyC}}^{\mathrm{inc}}$ of the former. In this case, although the Ly$\alpha$ efficiency of the faster I-front is lower (see Fig. \ref{fig:pspace_lya_eff_fd}), it may nonetheless produce a larger flux of Ly$\alpha$ photons because $F_{\mathrm{LyC}}^{\mathrm{inc}}$ is a factor of ten larger. For concreteness, consider the case with $\alpha = 1.5$ for which the efficiencies are $\zeta(10^3~\mathrm{km}~\mathrm{s}^{-1}) = 0.33$ and $\zeta(10^4~\mathrm{km}~\mathrm{s}^{-1}) = 0.19$. Note that, despite the lower efficiency, $F_{\mlya}^\mathrm{emit}$ is still a factor of $5.6$ larger for the faster moving I-front. This is just a consequence of the larger number of ionizing photons incident on the back of the faster moving I-front. More ionizing photons will produce more free electrons, thus more collisionally excited neutrals.  Note, however that the degree to which the $F_{\mathrm{LyC}}^{\mathrm{inc}}$ factor ``wins'' over $\zeta$ depends on $\alpha$. Repeating this example, but for $\alpha = 2.5$, we find that the \lya\ flux from the fast \IF\ is still greater by a factor of 2.7, but less so compared to the $\alpha=1.5$ case.  This discussion highlights some subtleties in interpreting Figure \ref{fig:pspace_lya_eff_fd}. We emphasize, however, the virtue of presenting our results in terms of $\zeta$: the dependence of this quantity on $\langle n_{\rm H}\rangle$ is completely encapsulated in $v_{\rm IF}$, which would not be true for $F_{\mlya}^\mathrm{emit}$.

Consider now the case of fixed $F_{\mathrm{LyC}}^{\mathrm{inc}}$ and $\alpha$. With fixed incident flux, a slower I-front can only result from moving through a higher local density of hydrogen (recall $v_{\rm IF} = F_{\mathrm{LyC}}^{\mathrm{inc}}/\langle n_{\rm H} \rangle$). In this case the Ly$\alpha$ flux just increases in proportion to $\zeta$, so a slower I-front produces a larger $F_{\mlya}^\mathrm{emit}$. Physically, this owes to the larger number of neutrals present for collisional excitation when $\langle n_{\rm H} \rangle$ is larger, and the fact that there is more cooling inside of a slower moving I-front.  As a last example, consider now fixed $v_{\rm IF}$ and $F_{\mathrm{LyC}}^{\mathrm{inc}}$ (so fixed $\langle n_{\rm H} \rangle$), but allow $\alpha$ to vary. A smaller $\alpha$ corresponds to harder spectra for which the average photon energy is larger. In this case, a harder spectrum (moving downward in Fig. \ref{fig:pspace_lya_eff_fd}) increases $F_{\mlya}^\mathrm{emit}$ because the more energetic ionizing photons make the I-front wider and its internal temperatures hotter, which boosts the efficiency of collisional excitation cooling. It follows from this overall discussion that the flux of \lya\ photons emitted from an I-front is largest if 3 conditions are met simultaneously: (1) the incident ionizing flux is large; (2) the incident spectrum is hard; (3) the \IF\ is traveling through a cosmological over-density, which causes it to propagate more slowly. 

Lastly, we provide an interpolation table for Figure \ref{fig:pspace_lya_eff_fd} in a Github repository\footnote{\href{https://github.com/bayu-wilson/1DRT_quantify_reion_fronts/tree/main/results/interp_tables/fd_parameter_space_interp_table.txt}{https://github.com/bayu-wilson/1DRT\_quantify\_reion\_fronts/tree/main/results/interp\_tables}}, which may be readily applied to reionization simulations that track $\alpha$ and $v_{\rm IF}$.  In most simulation frameworks, however, it is considerably more difficult to track $\alpha$, which requires multi-frequency RT with enough frequency bins and spatial resolution to reliably model spectral hardening by intergalactic absorptions.  In this case our results can still be of use under reasonable assumptions about $\alpha$, as long as $v_{\rm IF}$ can be calculated in the simulation.  For added convenience, we also provide a quartic fit to the \lya\ efficiency at fixed $\alpha_\mathrm{IF} = 1.5$, which is representative of expectations from stellar population synthesis models (see discussion in \cite{D_Aloisio_2019}) as well as  $\alpha_\mathrm{IF} = 0.85$, motivated by our estimates of spectral hardening by the IGM in Appendix \ref{appendix:hardening}:
\begin{equation}\label{eq:cubic_fit}
    \zeta(\alpha,v_\mathrm{IF}) = \frac{F^{\mathrm{emit}}_{\mathrm{Ly}\alpha}}{F^{\mathrm{inc}}_{\mathrm{ion}}} = \sum_{i=0}^{4} A_n \log_{10}\left(\frac{v_\mathrm{IF}}{[\mathrm{km/s}]}\right)^n,
\end{equation}
where $(A_0,A_1,A_2,A_3,A_4) = (1.7677$, $-1.7873$, $8.7365\times 10^{-1}$, $-1.8888\times 10^{-1}$, $1.4368\times 10^{-2} )$ for $\alpha=1.5$ and $(A_0,A_1,A_2,A_3,A_4) = ( 1.7679$,$-1.6657$, $8.3349\times10^{-1}$, $-1.7765\times10^{-1}$, $1.2834\times10^{-2})$ for $\alpha=0.85$.
These fits are accurate to $<2\%$ everywhere within $300 \lesssim v_{\rm IF} \lesssim 5 \times 10^4$ km/s.



\section{Conclusion}
\label{sect:conclusion}

In this paper we quantified Ly$\alpha$ emissions by I-fronts during reionization. The central quantity that we used was the dimensionless Ly$\alpha$ production efficiency, $\zeta$, or the ratio of emitted Ly$\alpha$ number flux to the flux ionizing photons being absorbed by the I-front.   We have argued that the likely parameter space of $\zeta$ during reionization can to a good approximation be covered by two variables: the I-front speed and the spectral index of the incident ionizing radiation.  We found that the main effect of small-scale density fluctuations within the I-fronts is to add scatter to $\zeta$; the fluctuations do not boost the mean efficiency appreciably.

The main results of this paper are presented in Figure \ref{fig:pspace_lya_eff_fd}, which lays out the parameter space of $\zeta$.     We found that the Ly$\alpha$ emissions are most intense if three conditions are met simultaneously:  (1) the ionizing flux driving the I-front is large; (2) the spectrum of this radiation is hard; (3) the I-front is traveling more slowly
through an over-density. The main application of our results are that they provide a sub-grid model for Ly$\alpha$ emissions in lower-resolution simulations of reionization, which do not spatially or temporally resolve the I-front radiative processes.  We make our results publicly available in a Github repository referenced below.  For simpler applications we also provide the fitting function in equation (\ref{eq:cubic_fit}), which assumes a single spectral index of $\alpha = 1.5$. 

Our results can be used to model the Ly$\alpha$ intensity signal from reionization and forecast its detectability in future intensity mapping surveys, a topic taken up independently by Refs. \cite{yang2023lymanalpha,koivu2023lymanalpha}.  The main motivation for our work, however, is the growing consensus that reionization ended below $z=6$, perhaps as late as $z\approx 5.3$, and that signatures of reionization's last phases may be present in existing $z\sim 6$ quasar absorption spectra \cite[e.g.][]{2015MNRAS.447.3402B, Kulkarni_2019, 2020MNRAS.491.1736K, Nasir_2020, 2022MNRAS.514...55B, 2022ApJ...932...76Z}. If this picture is correct, then long dark gaps in the coeval Ly$\alpha$ and Ly$\beta$ forests at $z>5.5$ could be signposts for locating some of the last neutral islands, providing targets for deep surveys aimed at detecting their spatially extended, low surface-brightness I-front emissions.  In Paper II, we present detailed models of neutral island emissions and we discuss the survey parameters required to detect them.

\acknowledgments
The authors thank Matt McQuinn for providing the cosmological hydrodynamics skewers used in this work, and for helpful comments on this manuscript. We also thank Zheng Zheng for helpful discussions. A.D.’s group was supported by NASA 19-ATP19-0191, NSF AST-2045600, and JWSTAR-02608.001-A.  E.V. is supported by NSF grant AST-2009309 and NASA ATP grant 80NSSC22K0629. Computations were performed under NSF ACCESS allocations TG-PHY210041 and TG-PHY230063. Expanse is an NSF-funded system operated by the San Diego Supercomputer Center at UC San Diego, and is available through the ACCESS program.

\bibliographystyle{JHEP}
\bibliography{references}

\appendix 
\section{Numerical Convergence}
\label{App:convergence}

\begin{figure*}
    \centering
    \includegraphics[width=\textwidth]{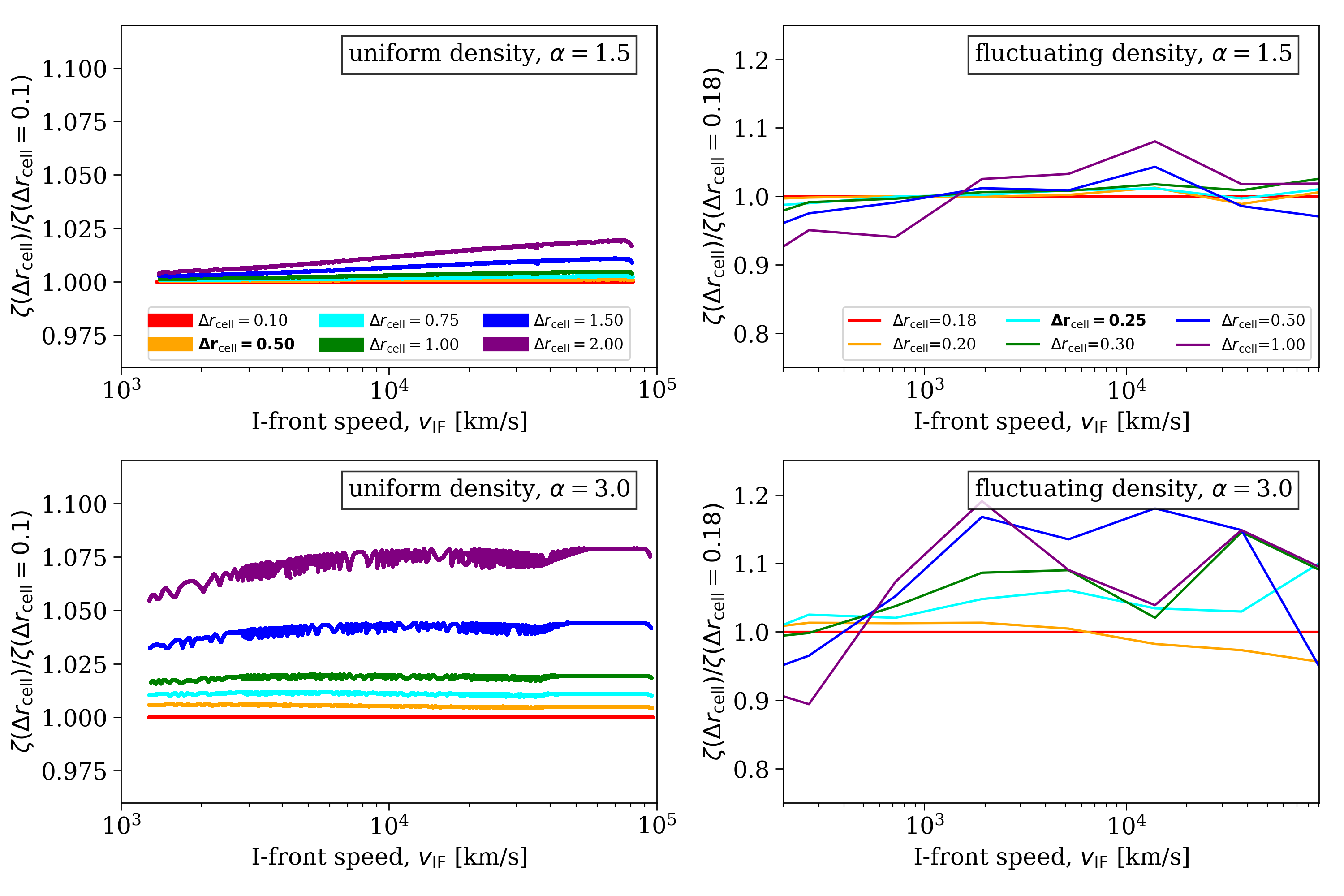}
    \caption{Numerical convergence of the \lya\ production efficiency calculation in our \odrt\ simulations. The left and right panels correspond to uniform- and fluctuating-density runs, respectively.  All results are shown as ratios with respect to the test run with the highest resolution.  The cases with bold fonts in the plot legends correspond to the resolution of our production runs.  The top rows assume an incident ionizing spectra index of $\alpha =1.5$, while the bottom row assumes $\alpha = 3$.  Note that the latter has the most stringent convergence requirements among our production runs because a softer spectrum generally leads to sharper I-fronts.  }
    \label{fig:csc}
\end{figure*}

Here we present numerical convergence tests of our \odrt\ simulations. In Figure \ref{fig:csc} we show a suite of tests in which we vary RT cell size, according to the colors in the plot legend.  We show the Ly$\alpha$ production efficiency, $\zeta$, normalized by the highest resolution run among our tests.  The bold fonts in the plot legend denote the resolutions of our production runs in the main text.  The left and right panels show tests of our uniform- and fluctuating-density runs, respectively.  In the top panels we assume fixed $\alpha = 1.5$, while in the bottom panels we show results for $\alpha =3$.  Note that the latter has the most stringent resolution requirements among our production runs because a softer spectrum results in sharper I-fronts, i.e. the I-fronts are less spatially extended.  

The left panels indicate that $\zeta$ is converged to within $2\%$ across the range of $v_{\rm IF}$ in our uniform-density runs for RT grid cell sizes $\Delta r_{\rm cell}\leq 1$ pkpc. (Our uniform-density production runs use $\Delta r_{\rm cell}\leq 0.5$ pkpc.) 
In the convergence test for fluctuating-density runs (right panels), we randomly chose 5 skewers extracted from the same hydro simulation described in the main text and assigned them different source luminosities to cover the $v_{\rm IF}$ range. For each skewer, we compute the mean $\zeta$ in logarithmically spaced bins of $v_{\rm IF}$. The smallest cell size in our fluctuating-density runs is limited by the spatial resolution of the hydro skewers, $\Delta r_{\rm cell} =  0.18$ pkpc, so we normalize all the curves in the right panels by this value. We see that the efficiency is converged to within a $<10$ percent for our production run resolution of $\Delta r_{\rm cell}\leq 0.25$ pkpc.   

\section{The spectra of ionizing radiation incident on intergalactic I-fronts}
\label{app:spectra}

\subsection{On the validity of the power-law approximation}

In this work, we parameterize the spectrum of the ionizing radiation incident on I-fronts using an effective spectral index, $\alpha_\mathrm{eff}$. Even if the source spectrum were a perfect power-law, spectral filtering preferentially absorbs ionizing radiation closer to $1$ Ry, which could lead to a significant deviation from the power-law assumption. Here we explore how well the power-law approximation holds for typical conditions during reionization.

We apply the \odrt\ code on 76 skewers extracted from the same cosmological hydrodynamics simulation that was used in our production runs. Similarly, these density fields are re-scaled from $z=7.1$ to $z=5.7$. To increase the path-length  of our RT calculations, we randomly splice together two skewers, yielding a total of 38 sight lines, each with a length $25$ h$^{-1}$cMpc. A plane-parallel source of radiation with spectral index $\alpha = 1.5$ is placed at the origin and then the RT equation is solved at each RT time step until the \IF\ has propagated through the entire skewer.\footnote{Note, in main text, we applied a strong homogeneous ionizing background to the gas behind the I-front as a method to isolate the incident spectral index.  However, we turn off this feature and run the code in ``standard'' mode for the calculations in this appendix. Additionally note that, in this calculation, we disable the thermal evolution feature of the \odrt\ code and use a constant temperature of $10^4$ K. This significantly decreases the computational load.} As before, we use 30 frequency bins for the RT calculation.  The source luminosity is calibrated such that the \HI\ photoionization rate in the ionized gas behind the \IF, averaged over all skewers, approximately matches $z=5.7$ measurements of Ref. \cite{2023arXiv230402038G}.  (Figure \ref{fig:GammaHI_alpha} in the next section shows the mean $\Gamma_{\rm HI}$ from our skewers as a function of distance from the source.)

\begin{figure}
    \centering
    \includegraphics[width=\textwidth]{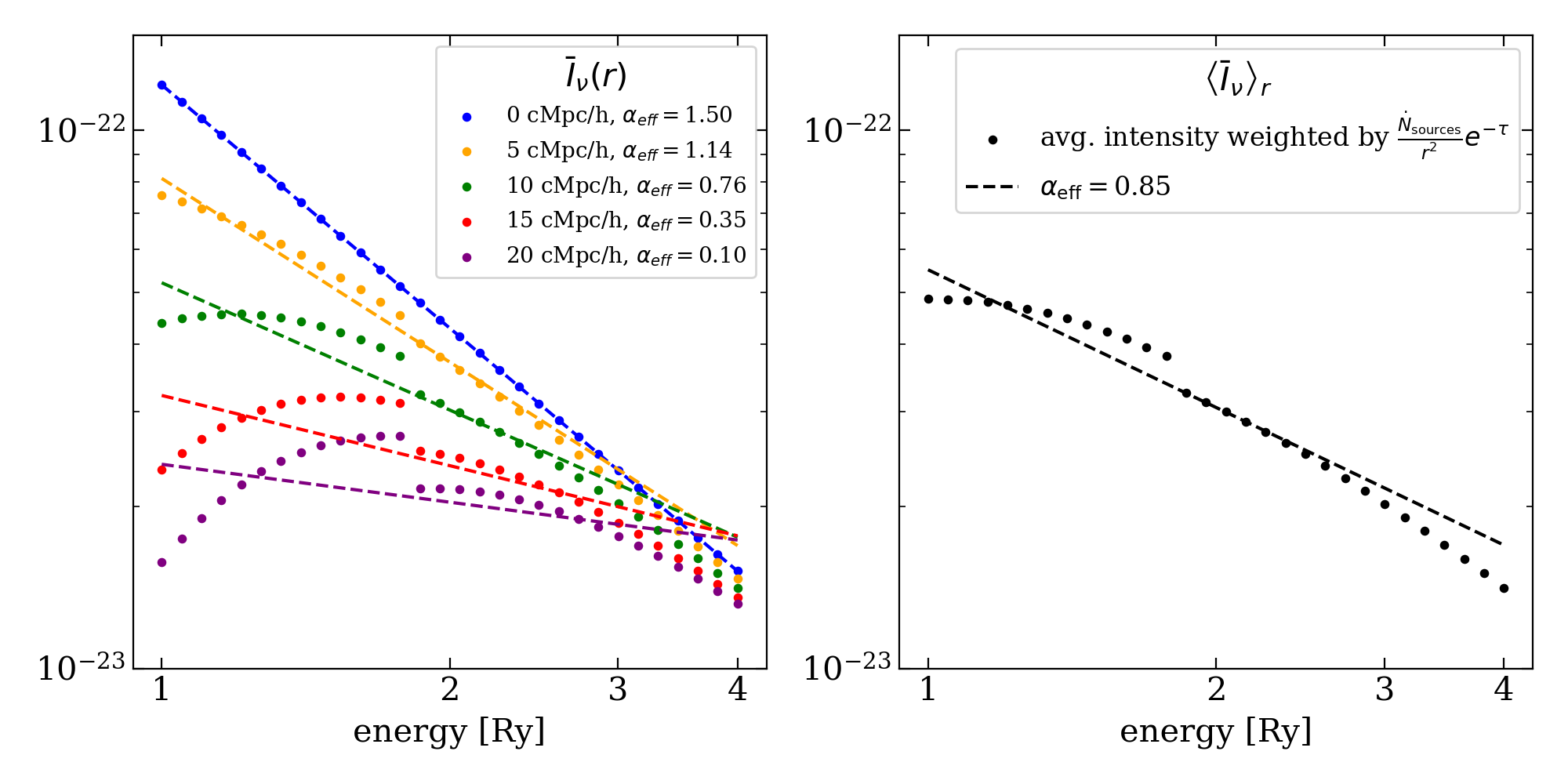}
    \caption{Testing how well a power-law approximates the spectrum of ionizing radiation incident on I-fronts during reionization. \textit{Left}: Changes in the spectra shape at different distances from a power-law point source with intrinsic spectral index $\alpha = 1.5$. Data points show the specific intensity [erg s$^{-1}$ cm$^{-2}$ sr$^{-1}$ Hz$^{-1}$], averaged over 38 sight lines through a cosmological hydro simulation sight lines from a 1D-RT. Five different distances from $0-25$ h$^{-1}$ cMpc are shown, as indicated in the plot legend. The dashed lines show power-law approximations with effective spectral indices indicated in the legend (see main text). For a single source, significant deviations from a power-law begin to occur at distances $\gtrsim 5~h^{-1}$cMpc from the source. \textit{Right}: More realistically, the ionizing spectrum hitting an I-front will be averaged over many sources at different distances. Here we attempt to model this effect using the source distribution from a cosmological simulation.  The data points show the result after averaging over the simulated source population, where each source's contribution to the spectrum is weighted by the rate at which it delivers ionizing photons to the I-front (see main text). The dashed line shows a power-law with effective index $\alpha_\mathrm{eff} = 0.85$. From this test, we conclude that, if a power-law is a reasonable approximation to the intrinsic spectrum of the sources, then a power-law (with hardened spectral index) likely remains a reasonable first approximation to the spectral shape of the radiation driving intergalactic I-fronts, in spite of filtering by the intervening IGM.}
    \label{fig:power_law_tests}
\end{figure}

We find the specific intensity in every spatial bin, $\bar I_\nu(r)$, averaged over all 38 skewers. (Note that the over-bar here and throughout this appendix denotes an average over skewers.)  In the left panel of Figure \ref{fig:power_law_tests}, the data points show $\bar I_\nu(r)$ at 5  travel distances through an ionized IGM from $0-20$ h$^{-1}$cMpc, as indicated in the plot legend. The kink in the spectra near $1.8$ Ry is due to  \HeI\ ionizations. The dashed lines show power-law spectra with effective spectral indices, $I_\nu \propto \nu^{-\alpha_\mathrm{eff}}$, estimated as follows. Using the average spectrum, the frequency-averaged \HI\ ionizing cross section is,
\begin{align}
    \left<\sigma_\mathrm{HI}\right>_\nu &= \frac{\int_{\nu_0}^{4\nu_0}  \sigma_\mathrm{HI}(\nu_j) \frac{\bar{I}_\nu}{h \nu} d\nu}{\int_{\nu_0}^{4\nu_0}  \frac{\bar{I}_\nu}{h \nu}d\nu} \rightarrow
     \frac{\sum_{j=0}^{N_\nu} \sigma_\mathrm{HI}(\nu)\bar{n}_j }{\bar{N}_\nu}.
     \label{eq:alphaeff_1}
\end{align}
The limits of integration are from $\nu_0$ to $4\nu_0$ which corresponds to 1 to 4 Ry, $\bar{N}_\nu$ is the skewer-averaged number of photons in all frequency bins, and $\bar{n}_j$ is the skewer-averaged number of photons in the $j$th frequency bin. On the other hand, for a power-law spectrum with spectral index $\alpha$ between 1 and 4 Ry, the average cross section is,
\begin{equation}
    \left<\sigma_\mathrm{HI}\right>_\nu^\mathrm{analytic} = \sigma_\mathrm{HI}^0\frac{\alpha}{\alpha+\beta}\frac{1-4^{-\alpha-\beta}}{1-4^{-\alpha}},
    \label{eq:alphaeff_2}
\end{equation}
where the H photoionization cross section is assumed to have the frequency dependence $\sigma_{\rm HI}(\nu) = \sigma^0_{\rm HI} (\nu/\nu_0)^{-\beta}$. Here we will adopt $\beta = 2.75$, which is a good approximation for the frequencies of interest here \cite{1996ApJ...465..487V}.  We set the $\langle \sigma_{\rm HI} \rangle_\nu = \langle \sigma_{\rm HI} \rangle_\nu^{\rm analytic}$ and solve for the effective $\alpha$ of the simulated spectrum.

Figure \ref{fig:power_law_tests} shows that spectral filtering by the IGM has a small effect on the spectrum shape at distances $\lesssim 5~h^{-1}$cMpc. By a distance of $5~h^{-1}$cMpc, there is substantial spectral hardening, as indicated by the lower value of $\alpha_{\rm eff}$ (about this more below).  For larger distances, the shape begins to deviate significantly from the initial power-law form. These results are for a single source.  It is important to note that, however, that the ionizing background incident on any I-front during reionization will be an integrated contribution from many sources at a distribution of distances.

 To estimate this effect, we calculate the spatially-averaged intensity weighted by the rate of ionizing photons coming from different distances,
\begin{equation}
    \left<\bar I_\nu\right>_r = \frac{\int \bar I_\nu(r) g(r) dr}{\int g(r) dr}.
\end{equation}
where $g(r)=\frac{\dot N(r)}{r^2} e^{-\tau}$ is the weight function, $\dot N(r)$ is the rate of ionizing photons produced by a source at distance $r$ from the \IF, the factor of $r^{-2}$ accounts for geometric attenuation, and $\tau$ is an estimate of the \HI\ optical depth.

We estimate $g(r)$ using source catalogs derived from an Eulerian hydro simulation described in Paper II which we will only briefly summarize here. The simulations were run with the hydro code of Ref. \cite{2004NewA....9..443T} in a cubic periodic volume with side $L=40$ h$^{-1}$cMpc and a uniform spatial grid with $N=400^3$ cells. The density fields and dark matter halos were tracked between $z=12-5.4$ and source catalogs were populated by abundance matching to observed UV luminosity functions as described in Ref. \cite{2024MNRAS.531.1951C}. We use the galaxy catalog from the $z=5.74$ snapshot to get the galaxy positions and $\dot N$'s. For the work in Paper II, we ran RT in post-processing on this simulation to generate an ionization field characteristic of the ending phases of reionization, when the mostly ionized IGM is punctuated by some last remaining ``neutral islands'' in the cosmic voids.  For the calculation here, we center the simulation box on a neutral island and use the location of the neutral island center, $r_\mathrm{NI}$, and source location, $r_s$, to estimate the source-\IF\ distance, $r=r_s-r_\mathrm{NI}$. Although clearly an approximation, this gives us a reasonable model for $g(r)$. We emphasize that each source is ultimately weighted according to the rate at which it is delivering ionizing photons to the I-front, which is encapsulated with $g(r)$.  We limit the source distance to be $0<r<20$ h$^{-1}$cMpc. In practice, the contribution from sources further than this is small.

In addition to geometric attenuation, we apply an approximate attenuation factor of $e^{-\tau}$ from LyC absorption, where
\begin{equation}
    \tau = \int_0^r \left<\kappa_\nu\right>_{\alpha=1.5} dr'.
\end{equation}
Here, $r$ is the distance from source to I-front, and $\left<\kappa_\nu\right>_{\alpha=1.5}$ is the frequency-averaged absorption coefficient for the intrinsic truncated power-law spectrum of our sources. To obtain $\left<\kappa_\nu\right>_{\alpha=1.5}$, we assume a Lyman limit mean free path of $\lambda_\mathrm{mfp,HI} = 1/\kappa_{912} = 16.6$ h$^{-1}$cMpc at $z=5.7$, consistent with recent observational measurements \cite{2023ApJ...955..115Z, 2023arXiv230402038G}. We then translate this to $\left<\kappa_\nu\right>_{\alpha=1.5}$ using eq. (12) of \cite{2024MNRAS.531.1951C}, with column density distribution slope of $\beta_{\rm N} = 1.6$, obtained from their Appendix B.

The data points in the right panel of Figure \ref{fig:power_law_tests} show the result of this exercise.  The dashed curve again shows a power-law spectrum with effective spectral index $\alpha_\mathrm{eff}=0.85$, obtained in the same way as above. There are obvious deviations from the intrinsic power-law shape of the source spectrum, and the effective spectral index is significantly harder than the source spectral index of $\alpha = 1.5$. The latter should not be a surprise, however; H absorptions are expected to harden the ionizing spectrum substantially -- a topic that we consider in the next section.  Importantly, if a power-law is a reasonable approximation to the intrinsic spectrum of the sources, then a power-law (with hardened spectral index) likely remains a reasonable first approximation to the spectral shape of the radiation driving intergalactic I-fronts, in spite of filtering by the intervening IGM. Indeed, Figure \ref{fig:power_law_tests} suggests that the average weighted intensity deviates from the effective power-law at the $\lesssim 20\%$ level.

\subsection{Spectral hardening}
\label{appendix:hardening}

\begin{figure}
    \centering
    \includegraphics[width=\textwidth]{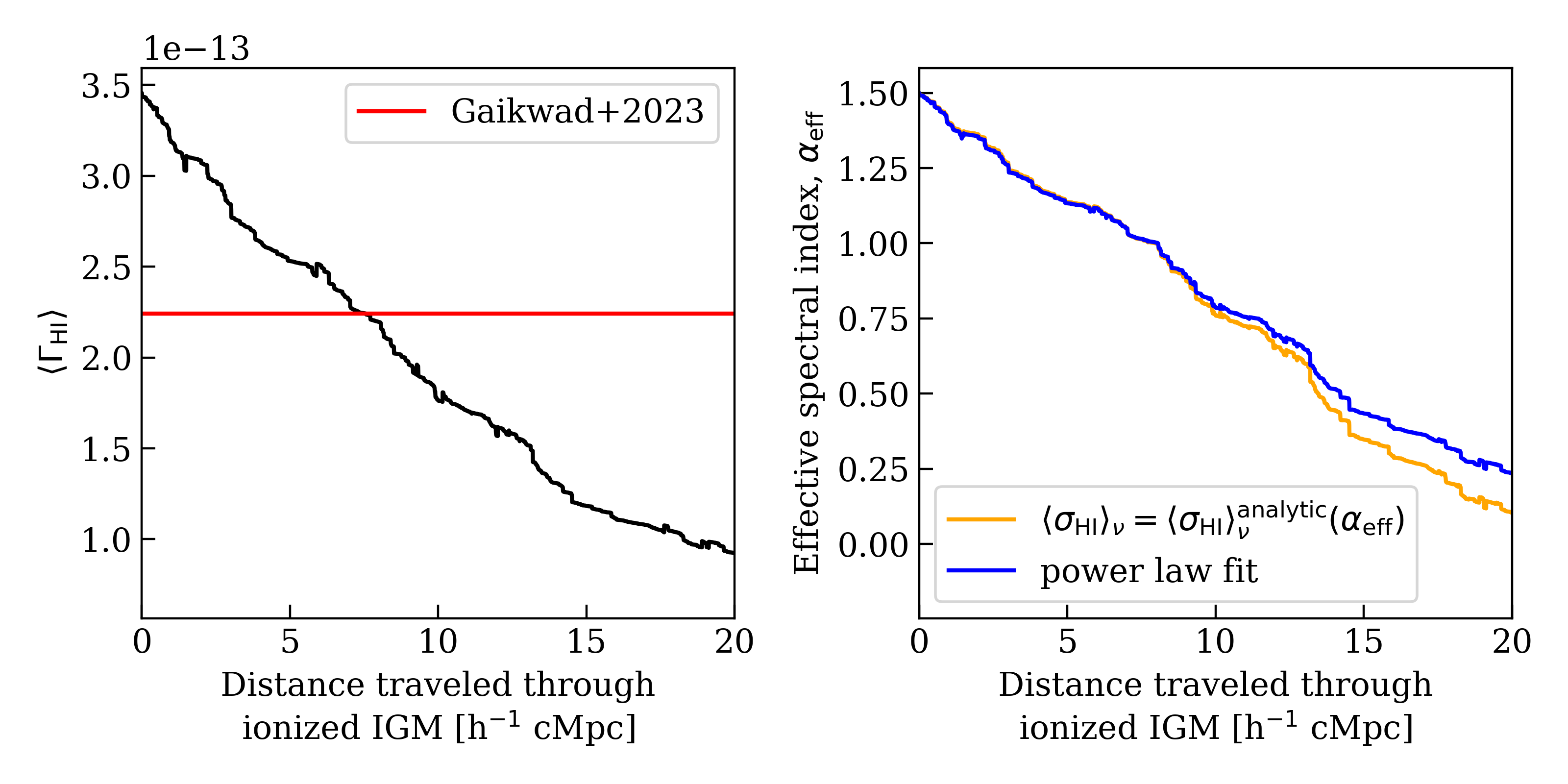}
    \caption{Spectral hardening by residual \HI\ absorption in the IGM at $z\sim 5.7$. We use the \odrt\ code from Paper I on 38 in-homogeneous density skewers from a cosmological hydrodynamics simulation. {\it Left:}  The \HI\ photo-ionization rate, $\Gamma_{\rm HI}$, averaged over the skewers, as a function of distance from a source.  The source luminosity was calibrated such that the average $\Gamma_{\rm HI}$ agrees with the measurement of \cite{2023arXiv230402038G} at $z=5.7$. {\it Right:}  The effective spectral index, $\alpha_{\rm eff}$, of ionizing radiation as a function of distance from the source measured in two different ways (see text). These results demonstrate that significant spectral hardening occurs as ionizing radiation traverses large distances through the IGM.}
    \label{fig:GammaHI_alpha}
\end{figure}

The main text shows that the spectral shape of the incident ionizing spectrum plays a significant role in setting the amount of \lya\ photons emitted from I-fronts. In addition to the properties of the sources, the incident spectral index depends on the amount of filtering by H absorption as the radiation traverses long distances through the IGM. Here we provide an estimation for the expected amount of spectral filtering by the IGM. 

We use the same 1D-RT setup that is described in the last appendix. The radiation spectrum at each cell is averaged over the 38 sight lines, producing a mean spectrum as a function of distance from the source, $\bar{I}_\nu$.   We then define an effective spectral index using the procedure described in the last section (see eqs. \ref{eq:alphaeff_1} and \ref{eq:alphaeff_2} above). For reference, the left panel of Figure \ref{fig:GammaHI_alpha} shows the mean $\Gamma_{\rm HI}$ from our skewers as a function of distance from the source. The red horizontal line represents the central value of the recent measurement by Ref. \cite{2023arXiv230402038G} from Ly$\alpha$ forest observations, $\Gamma_{\rm HI} = 0.224^{+0.223}_{-0.112} \times 10^{-12}$ s$^{-1}$.  (Note that the 1-$\sigma$ error bars are not shown because they are out of range of the plot.) The purpose of this panel is to show that our 1D-RT setups have been calibrated to reproduce the ionization state of the IGM to within existing empirical bounds. 

The orange curve in the right panel of Figure \ref{fig:GammaHI_alpha} shows the effective $\alpha$ measured from our \odrt\ simulations at various distances traveled through an in-homogeneous ionized IGM. We find that significant spectral hardening can occur when traveling large distances through an ionized medium. As an alternative way of defining the effective spectral index, the blue curve shows the effective index obtained by fitting a truncated power law, $I_\nu \propto \nu^{-\alpha_\mathrm{eff}}$, to the stacked simulated spectra at each distance.  Encouragingly, we find that $\alpha_{\rm eff}$ derived in this manner matches that from our fiducial method (eqs. \ref{eq:alphaeff_1} and \ref{eq:alphaeff_2}) to within $20\%$ up until the \IF\ has traveled $15~h^{-1}$ cMpc. This supports the conclusion drawn in the last section.  The right panel of Figure \ref{fig:GammaHI_alpha} also gives us a handle on how much the IGM hardens a spectrum. A galaxy spectrum with emitted spectral index $\alpha_{*}=1.5$ could plausibly harden to $\alpha=0.5$ after traveling $10-15~h^{-1}$cMpc through the IGM. These results motivate the wide range of incident spectral indices explored in Figure \ref{fig:pspace_lya_eff_fd}.

\end{document}